\documentclass[12pt]{article}
\newcommand{\ba}{\begin{array}}
                 \newcommand{\ea}{\end{array}}
\newcommand{\nn}{\nonumber\\}
\newcommand{\R}{{\bf R}}

\newcommand{\del}{\partial}

\newcommand{\rar}{\rightarrow}

\newcommand{\dis}{\displaystyle}

\makeatletter
\@addtoreset{equation}{section}

\makeatother

\begin{document}

\begin{titlepage}
\null
\begin{flushright}
hep-th/0507112
\\
July, 2005
\end{flushright}

\vskip 1.5cm
\begin{center}

 {\Large \bf On Reductions of Noncommutative}

\vskip 0.5cm

{\Large \bf Anti-Self-Dual Yang-Mills Equations}

\vskip 1.7cm
\normalsize

{\large Masashi Hamanaka\footnote{The author visits Oxford 
from 16 August, 2005 to  15 August, 2006.\\
E-mail: hamanaka@math.nagoya-u.ac.jp, hamanaka@maths.ox.ac.uk}}

\vskip 1.5cm

        {\it Graduate School of Mathematics, Nagoya University,\\
                     Chikusa-ku, Nagoya, 464-8602, JAPAN}
\vskip 0.5cm

        {\it Mathematical Institute, University of Oxford,\\
                     24-29, St Giles', Oxford, OX1 3LB, UK}

\vskip 1.5cm

{\bf \large Abstract}

\end{center}

We show that
various noncommutative integrable equations
can be derived from noncommutative anti-self-dual 
Yang-Mills equations in the split signature,
which include noncommutative versions of 
Korteweg-de Vries, Non-Linear Schr\"odinger,
$N$-wave, Davey-Stewartson and 
Kadomtsev-Petviashvili equations.
$U(1)$ part of gauge groups
for the original Yang-Mills equations play crucial roles 
in noncommutative extension of Mason-Sparling's 
celebrated discussion.
The present results would 
be strong evidences for noncommutative Ward's conjecture
and imply that these noncommutative integrable
equations could have the corresponding physical
pictures such as reduced configurations
of D0-D4 brane systems in open N=2 string theories.  
Possible applications to 
the D-brane dynamics are also discussed.

\end{titlepage}
\clearpage
\baselineskip 5.8mm

\section{Introduction}

Field theories on non-commutative (NC) spaces have been studied
intensively for the last several years. NC gauge theories are
equivalent to ordinary gauge theories in the presence of
background magnetic fields and succeeded in revealing various
aspects of them \cite{NC}. NC solitons especially play
important roles in the study of D-brane dynamics, such as the
confirmation of Sen's conjecture on tachyon condensation 
\cite{Sen}. One of the distinguished
features of NC theories is resolution of singularities. This gives
rise to various new physical objects such as U(1) instantons and
makes it possible to analyze singular configurations 
as usual \cite{YM}.

NC extension of integrable equations is
also one of the hot topics
(For reviews, see \cite{Integ}.)
and has been 
studied from various viewpoints
by many authors \cite{CFZ}-\cite{ChLe}. 
These equations imply no gauge field
and NC extension of them perhaps might have no physical picture or
no good property on integrabilities. To make matters worse, NC
extension of $(1+1)$-dimensional equations introduces infinite
number of time derivatives, which makes it hard to discuss or
define the integrability. 

Nevertheless, some of them actually possess
integrable-like properties, such as the existence of infinite number of
conserved quantities and exact multi-soliton solutions,
and the linearizability and so on. 
Furthermore, a few of them can be derived from NC (anti-)self-dual
Yang-Mills (ASDYM) equations by suitable reductions,
which is just an example of NC version \cite{HaTo}
of Ward's conjecture \cite{Ward}, that is,
{\it almost all NC integrable equations are reductions 
of NC ASDYM equations}.
(For commutative discussions, see e.g.
\cite{MaWo}-\cite{IvPo}.)
This makes it possible to
give physical pictures into
lower-dimensional integrable equations and 
to apply analysis of NC solitons to that of the corresponding
D-branes. Origin of the integrable-like properties would be also
explained from the viewpoints of NC twistor theory and 
the preserved supersymmetry in the D-brane systems.
That is why confirmation of NC Ward conjecture 
is very important.

In this letter, we show that
various NC integrable equations\footnote{In the present letter, 
integrability of NC equations means 
existence of infinite number of conserved quantities or
exact multi-soliton solutions.}
can be derived from NC ASDYM equations
on $(2+2)$-dimensional flat spaces.
The signature of metric is $(+,+,-,-)$
and called the {\it split signature}.
In section 3, we extend Mason-Sparling's 
celebrated results \cite{MaSp} to NC spaces and 
derive NC Korteweg-de Vries (KdV) 
and Non-Linear Schr\"odinger (NLS)
equations on NC $(1+1)$-dimensional space-time.
We can prove that the gauge groups in original gauge
theories must be $G=GL(2,\R)$ not $G=SL(2,\R)$ 
and $G=U(2)$ not $G=SU(2)$. 
Hence we conclude that $U(1)$ part of the gauge groups
play crucial roles as usual in NC gauge theories.
In section 4, we derive NC 
$N$-wave, Davey-Stewartson (DS) and 
Kadomtsev-Petviashvili (KP)
equations on NC $(2+1)$-dimensional space-time
where the Lie algebras are operator algebras.
The present results imply that these NC integrable
equations could have actually the corresponding physical
pictures such as reduced D-brane configurations
of D0-D4 brane systems in open N=2 string theories
in background of $B$-fields \cite{LPS2}. 
Possible applications to the D-brane dynamics 
are also discussed in section 5.

\section{NC ASDYM Equations}

NC spaces are defined
by the noncommutativity of the coordinates:
\begin{eqnarray}
\label{nc_coord}
[x^i,x^j]=i\theta^{ij},
\end{eqnarray}
where $\theta^{ij}$ are real constants and
called the {\it NC parameters}.

NC field theories are given by the exchange of ordinary products
in the commutative field theories for the star-products and
realized as deformed theories from the commutative ones. 

The star-product is defined for ordinary fields on commutative
spaces. For Euclidean spaces, it is explicitly given by
\begin{eqnarray}
f\star g(x)&:=&\mbox{exp }
\left(\frac{i}{2}\theta^{ij} \partial^{(x^{\prime})}_i
\partial^{(x^{\prime\prime})}_j \right)
f(x^\prime)g(x^{\prime\prime})\Big{\vert}_{x^{\prime}
=x^{\prime\prime}=x}\nonumber\\
&=&f(x)g(x)+\frac{i}{2}\theta^{ij}\partial_i f(x)\partial_j g(x)
+O (\theta^2),
\label{star}
\end{eqnarray}
where $\del_i^{(x^\prime)}:=\del/\del x^{\prime i}$
and so on.
This explicit representation is known
as the {\it Moyal product} \cite{Moyal}.
The star-product has associativity:
$f\star(g\star h)=(f\star g)\star h$,
and returns back to the ordinary product
in the commutative limit:  $\theta^{ij}\rar 0$.
The modification of the product  makes the ordinary
spatial coordinate ``noncommutative,''
that is, $[x^i,x^j]_\star:=x^i\star x^j-x^j\star x^i=i\theta^{ij}$.

We note that the fields themselves take c-number values
as usual and the differentiation and the integration for them
are well-defined as usual.
A nontrivial point is that
NC field equations contain infinite number of
derivatives in general. Hence the integrability of the equations
are not so trivial as commutative cases.
Nevertheless, NC ASDYM equations are integrable in some sense.
(See, e.g. \cite{KKO,Takasaki,Hannabuss}.)

\vspace{3mm}

Let us consider Yang-Mills theories on 
$(2+2)$-dimensional NC spaces.
Here, we follow the convention in \cite{MaWo}.
NC ASDYM equation is derived from compatibility condition
of the following linear system:
\begin{eqnarray}
 (D_w-\zeta D_{\tilde{z}})\star \Psi = 0,~~~
 (D_z-\zeta D_{\tilde{w}})\star \Psi = 0,
\label{lin_asdym}
\end{eqnarray}
where $(z,\tilde{z},w,\tilde{w})$ and 
$D_z,D_w,D_{\tilde{z}},D_{\tilde{w}}$
denote the coordinates
of the original $(2+2)$-dimensional space and
covariant derivatives in the Yang-Mills theory, respectively.
Explicit representations are as follows:
\begin{eqnarray}
 &&\del_{w} A_z -\del_{z} A_w+[A_w,A_z]_\star =0,~~~
 \del_{\tilde{w}} A_{\tilde{z}} -\del_{\tilde{z}} A_{\tilde{w}}
 +[A_{\tilde{w}},A_{\tilde{z}}]_\star =0,\nn
 &&\del_{z} A_{\tilde{z}} -\del_{\tilde{z}} A_{z}
 +\del_{\tilde{w}} A_{w} -\del_{w} A_{\tilde{w}}
 +[A_z,A_{\tilde{z}}]_\star
 -[A_{w},A_{\tilde{w}}]_\star=0.
\label{asdym}
\end{eqnarray}
where $A_z,A_w,A_{\tilde{z}},A_{\tilde{w}}$
denote the gauge fields, respectively.
We note that the commutator part should remain
even when the gauge group is $U(1)$ because the elements of the
gauge group could be operators and the gauge group could be
considered to be non-abelian: $U(\infty)$.
This $U(1)$ part actually plays important roles
as NC gauge theories and gives rise to new physical objects 
\cite{NC, YM}.

Taking various reduction conditions of NC ASDYM equations
for, such as, choice of gauge groups,
translation invariance, gauge fixings, 
and further constraints on gauge fields,
we can get NC integrable equations
in $(2+1)$ and $(1+1)$-dimensional spaces.

\section{NC KdV and NLS Equations from ASDYM}

In this section, we discuss reductions of NC ASDYM equations
to $(1+1)$-dimensional space-time, 
including NC NLS eq., NC KdV equation,
which is NC extension of the Mason-Sparling's result \cite{MaSp}.
Relation to hyperCR Einstein-Weyl
structure has been also discussed 
by Dunajski and Sparling \cite{DuSp}. 
In this section, we follow the convention in \cite{MaWo}.

Let us consider the following NC ASDYM equation with
which is dimensionally reduced to 2-dimensional space-time 
with the coordinate $(t,x)=(z,w+\tilde{w})$:
\begin{eqnarray}
Q^\prime=0,~~~
 \dot{Q} + A_{w}^\prime
 +[A_{z},Q]_\star =0,~~~
A_{z}^\prime -\dot{A}_{w}
 +[A_{w},A_{z}]_\star=0.
\label{asdym_kdv}
\end{eqnarray}
where $Q, A_{w}$ and $A_{z}$ are $2\times 2$ matrices
and denote the original gauge fields,
and $Q^\prime:=\del Q/\del x,~ 
\dot{Q}:=\del Q/\del t$.

\begin{itemize}

\item NC KdV equation

Now let us take further reduction
on the gauge fields in the ASDYM equation (\ref{asdym_kdv})
as follows:
\begin{eqnarray}
&& Q=\left(\begin{array}{cc}0&0\\1&0\end{array}\right),~
 A_{w}=\left(\begin{array}{cc}q&~-1\\q^\prime+q\star q
&~-q\end{array}\right),\\
&& A_{z}=\left(\begin{array}{cc}
\displaystyle\frac{1}{2}q^{\prime\prime}
+aq\star q^\prime+(1-a)q^\prime\star q&-
q^\prime\\\phi&
-\displaystyle\frac{1}{2}q^{\prime\prime}-bq^\prime\star q-(1-b)
q\star q^\prime
\end{array}\right),\nonumber
\end{eqnarray}
where $a$ and $b$ are constants,
and $\phi$ is a differential polynomial of $q$.

The first equation of (\ref{asdym_kdv}) is trivially satisfied.
The second equation of (\ref{asdym_kdv}) becomes
\begin{eqnarray*}
 \left(\begin{array}{cc}0&~0\\(a-b)[q^\prime,q]_\star&~0\end{array}\right)=0,
\end{eqnarray*}
which leads to $a=b$. Then, 
the third equation of (\ref{asdym_kdv}) yields
\begin{eqnarray}
\phi&=&\displaystyle\frac{1}{2}q^{\prime\prime\prime}
+2q^{\prime}\star q^{\prime}
-\dot{q}
+a\left\{q\star q,q^\prime\right\}_\star
+(1-2a)q\star q^\prime \star q
\nn
&&+\frac{1}{2}\left\{q,q^{\prime\prime}\right\}_\star
+a[q, q^{\prime\prime}]_\star,
\label{red_kdv}
\\
\phi&=&
\displaystyle\frac{1}{2}q^{\prime\prime\prime}
+2q^{\prime}\star q^{\prime}
-\dot{q}
+a\left\{q\star q,q^\prime\right\}_\star
+(1-2a)q\star q^\prime \star q
\nn
&&
+\frac{1}{2}\left\{q,q^{\prime\prime}\right\}_\star
-a[q, q^{\prime\prime}]_\star,
\label{red_kdv2}
\\
\phi^\prime &=&\dot{q}^\prime +\dot{q}\star q +q\star \dot{q}
+\left\{\phi,q\right\}_\star
-\displaystyle\frac{1}{2}\left\{q^{\prime\prime},q\star q
+q^\prime\right\}_\star
-a\left\{q\star q\star q,q^\prime\right\}_\star\nn
&&-(1-a)\left\{q,q^{\prime}\star q^{\prime}
+q\star q^\prime\star q\right\}_\star
-2aq^\prime \star q\star q^\prime,
\label{red_kdv3}
\end{eqnarray}
where $\left\{A,B\right\}_\star :=A\star B +B\star A$.

{}From Eqs. (\ref{red_kdv}) and (\ref{red_kdv2}),
we get $a[q, q^{\prime\prime}]_\star=0$, hence, $a=0$.
(The published version shows incorrect derivation of $a=0$.)

If we take
\begin{eqnarray}
\phi=\displaystyle\frac{1}{4}q^{\prime\prime\prime}
+\frac{1}{2}q^{\prime}\star q^{\prime}
+\frac{1}{2}\left\{q,q^{\prime\prime}\right\}_\star
+q\star q^\prime \star q,
\end{eqnarray}
Eqs. (\ref{red_kdv}) and (\ref{red_kdv2}) 
become a NC potential KdV equation:
\begin{eqnarray}
 \dot{q}=\displaystyle\frac{1}{4}q^{\prime\prime\prime}
+\frac{3}{2}q^{\prime}\star q^{\prime}.
\label{pkdv}
\end{eqnarray}
This is derived from the NC KdV equation 
\begin{eqnarray}
 \dot{u}=\displaystyle\frac{1}{4}u^{\prime\prime\prime}
+\frac{3}{4}\left(u^{\prime}\star u+u\star u^{\prime}\right),
\label{kdv} 
\end{eqnarray}
by setting $2q^\prime =u$.
Then Eq. (\ref{red_kdv3}) 
also leads to the NC potential KdV equation as follows
\begin{eqnarray*}
\left(
-\dot{q}+\displaystyle\frac{1}{4}q^{\prime\prime\prime}
+\frac{3}{2}q^{\prime}\star q^{\prime}
\right)^\prime
+q\star
\left(
-\dot{q}+\displaystyle\frac{1}{4}q^{\prime\prime\prime}
+\frac{3}{2}q^{\prime}\star q^{\prime}
\right)
+\left(
-\dot{q}+\displaystyle\frac{1}{4}q^{\prime\prime\prime}
+\frac{3}{2}q^{\prime}\star q^{\prime}
\right)\star q=0.
\end{eqnarray*}

In summary, only when $a=b=0$,
we can obtain the NC KdV equation (\ref{kdv})
by reduction of NC ASDYM equation
under the reduction conditions:
\begin{eqnarray}
&& Q=\left(\begin{array}{cc}0&0\\1&0\end{array}\right),~
 A_{w}=\left(\begin{array}{cc}q&~-1\\q^\prime+q\star q
&~-q\end{array}\right),\\
&& A_{z}=\left(\begin{array}{cc}
\displaystyle\frac{1}{2}q^{\prime\prime}
+q^\prime\star q&-
q^\prime\\  
\displaystyle\frac{1}{4}q^{\prime\prime\prime}
+\frac{1}{2}q^{\prime}\star q^{\prime}
+\frac{1}{2}\left(q\star q^{\prime\prime}
+q^{\prime\prime}\star q\right)
+q\star q^\prime \star q
&
-\displaystyle\frac{1}{2}q^{\prime\prime}-q\star q^\prime
\end{array}\right),\nonumber
\end{eqnarray}

We note that the gauge group is not $SL(2,{\bf R})$ but $GL(2,{\bf R})$
on NC spaces because the matrix $A_{z}$ is not traceless.
This result also reflects the importance of
$U(1)$ part of the original gauge group.

\item NC NLS equation

Now let us take another further reduction
on the gauge fields in the ASDYM equation (\ref{asdym_kdv})
as follows:
\begin{eqnarray}
&& Q=\kappa\left(\begin{array}{cc}1&0\\0&-1\end{array}\right),~
 A_{w}=\left(\begin{array}{cc}0&\psi\\\tilde{\psi}&0\end{array}\right),\nn
&& A_{z}=\frac{1}{2\kappa}\left(\begin{array}{cc}
a\psi\star\tilde{\psi} +(1-a) \tilde{\psi}\star\psi
&\psi^\prime\\
-\tilde{\psi}^\prime&
-b\tilde{\psi}\star\psi-(1-b)\psi\star\tilde{\psi}
\end{array}\right).
\end{eqnarray}
These conditions automatically solve the first and the second equations
in (\ref{asdym_kdv}) and the third equation is reduced to
\begin{eqnarray*}
&&\frac{a-1}{2\kappa}\left(
\psi^\prime\star \tilde{\psi}
+\psi\star \tilde{\psi}^\prime
-\tilde{\psi}^\prime\star \psi
-\tilde{\psi}\star \psi^\prime\right)=0,\\
&&
\frac{1-b}{2\kappa}\left(
\tilde{\psi}^\prime\star \psi
+\tilde{\psi}\star \psi^\prime
-\psi^\prime\star \tilde{\psi}
-\psi\star \tilde{\psi}^\prime\right)=0,\\
&&
-\dot{\psi}+\frac{1}{2\kappa}\left(
\psi^{\prime\prime}
-(a+b)\psi\star \tilde{\psi}\star\psi
-(1-a)\tilde{\psi}\star \psi\star\psi
-(1-b) \psi\star\psi\star\tilde{\psi}
\right)=0,\\
&&
-\dot{\tilde{\psi}}+\frac{1}{2\kappa}\left(
-\tilde{\psi}^{\prime\prime}
+(a+b)\tilde{\psi}\star \psi\star\tilde{\psi}
+(1-a)\tilde{\psi}\star \tilde{\psi}\star\psi
+(1-b) \psi\star\tilde{\psi}\star\tilde{\psi}
\right)=0.
\end{eqnarray*}
We can also see that only when $a=b=1$,
these equations are reduced to
\begin{eqnarray}
 2\kappa \dot{\psi}=\psi^{\prime\prime}-2\psi\star\tilde{\psi}\star\psi,~~~
 2\kappa \dot{\tilde{\psi}}=-\tilde{\psi}^{\prime\prime}
+2\tilde{\psi}\star\psi\star\tilde{\psi}.
\end{eqnarray}
By taking $\dis \kappa=i/2, 
\tilde{\psi}=\bar{\psi}$, we get
the NC repulsive NLS equation with $G=U(1,1)$:
\begin{eqnarray}
 i\dot{\psi}=\psi^{\prime\prime}-2\psi \star \bar{\psi} \star \psi,
\end{eqnarray} 
and by taking $\dis \kappa=i/2, \tilde{\psi}=-\bar{\psi}$,
the NC attractive NLS equation \cite{DiMH2} with $G=U(2)$:
\begin{eqnarray}
 i\dot{\psi}=\psi^{\prime\prime}+2\psi \star \bar{\psi} \star \psi.
\end{eqnarray}

In summary, only when $a=b=1$,
we can obtain the NC NLS equation
by reduction of NC ASDYM equation
under the reduction conditions:
\begin{eqnarray}
Q=\frac{i}{2}\left(\begin{array}{cc}1&0\\0&-1\end{array}\right),~
A_{w}=\left(\begin{array}{cc}0&\psi\\
\varepsilon\bar{\psi}&0\end{array}\right),~
A_{z}=i\varepsilon\left(\begin{array}{cc}
-\psi\star\bar{\psi}
&-\varepsilon\psi^\prime\\
\bar{\psi}^\prime&
\bar{\psi}\star\psi
\end{array}\right),
\end{eqnarray}
where $\varepsilon=\pm 1$.
The case for $\varepsilon=-1$ coincides with 
results by Legar\'e \cite{Legare}.

We note that the gauge group must be $U(2)$ or $U(1,1)$
not $SU(2)$ or $SU(1,1)$, respectively on NC spaces 
because the matrix $A_{z}$ is not traceless. 
We prove that 
$U(1)$ part of the gauge group 
is crucial in this case also, and
the choice by Legar\'e \cite{Legare}
is unique for the reduction.

\end{itemize}

Here we would like to point out that
{\it every} NC conservation laws on $(1+1)$-dimensional 
space-time can be derived 
from NC ASDYM equations by a special reduction.\footnote{
The derivation in this paragraph is trivial,
because the gauge fields $A_z$ and $A_w$ in Eq. (\ref{triv})
can be set to zero via the following gauge transformation:
$A_{z,w}\rightarrow g^{-1}A_{w,z}g+g^{-1}\del_{z,w}g$,
where
$g=\left(\begin{array}{cc}1&0\\h&1\end{array}\right)$
and $h(t,x)=-\int^x\sigma(t,y)dy$.
This is an expected result.
(The author thanks L.~Mason for commenting on
this point, during stay at Oxford supported by the Yamada Science
Foundation.)}
Let us consider the following $G=SL(2,\R)$ 
NC ASDYM equation,
which is dimensionally reduced to 2-dimensional space-time 
with the coordinate $(t,x)=(z,w)$ and a fixed gauge
$A_{\tilde{z}}=A_{\tilde{w}}=0$:
\begin{eqnarray}
\del_{x} A_z -\del_{t} A_w+[A_w,A_z]_\star =0.
\end{eqnarray} 
By taking further reduction
\begin{eqnarray}
\label{triv}
A_z=\left(\begin{array}{cc}0&0\\J&0\end{array}\right),~
A_w=\left(\begin{array}{cc}0&0\\\sigma&0\end{array}\right),
\end{eqnarray}
we get an {\it arbitrary} conservation law
\begin{eqnarray}
 \del_t \sigma = \del_x J,
\end{eqnarray}
because $\sigma$ and $J$ can be taken  arbitrarily.
For example, $\sigma=4u,~J=u^{\prime\prime}+3u\star u$
yields the NC KdV equation. (cf. \cite{Legare}.)

\section{NC $N$-wave, DS and KP Equations from ASDYM}

In this section, we discuss reductions
of NC ASDYM equations where the components of
gauge fields are elements of
an infinite dimensional Lie algebra
of formal matrix differential operators in an auxiliary
variable $\del_y$:
\begin{eqnarray}
 g=\left\{a_0+a_1\del_y +a_2 \del_y^2\right\},
\end{eqnarray}
where $a_0,a_1,a_2$ belong to a ring of $N\times N$ 
matrix functions of $y$.
Here we follow the convention in \cite{ACT}.

\begin{itemize}

\item NC $N$-wave equation

Let us suppose that all gauge fields depend 
on $t=z$ and $x=w$ only, which is 
a simple dimensional reduction.
The linear system (\ref{lin_asdym}) is reduced
as follows :
\begin{eqnarray}
 \del_x \Psi=(A_w+\zeta A_{\tilde{z}})\star \Psi,~~~
 \del_t \Psi=(A_z+\zeta A_{\tilde{w}})\star \Psi,
\end{eqnarray}
where $\Psi$ is a function of $x,y,t$ and $\zeta$.
Now let us take further reduction condition as
$A_w=U+A\del_y,~A_z=V+B\del_y,~
A_{\tilde{z}}=A_{\tilde{w}}=0$, where
$A,B$ are constant, commuting $N\times N$
matrices: $[A,B]=0$. By taking $\Psi = \Phi e^{-\zeta y}$,
we get a reduced linear system:
\begin{eqnarray}
 \del_x \Phi=(U+A\del_y)\star \Phi,~~~
 \del_t \Phi=(V+B\del_y)\star \Phi.
\end{eqnarray}
The compatible condition for the reduced linear system
gives rise to a reduced ASDYM equation:
\begin{eqnarray}
U_t-V_x+[U,V]_\star +A\star V_y-B\star U_y=0,~~~
[A,V]_\star=[B,U]_\star,
\end{eqnarray}
where $U_t:=\del U/ \del t, V_x:=\del V/\del x$ etc. 
This results a non-linear equation:
\begin{eqnarray}
 [A,Q_t]_\star-[B,Q_x]_\star
 +[[A,Q]_\star,[B,Q]_\star]_\star
+A\star [B,Q_y]_\star-B\star [A,Q_y]_\star=0,
\end{eqnarray}
where $U=[A,Q]_\star,V=[B,Q]_\star$.

If we take $A, B$ as diagonal matrices:
$A_{ij}=a_i\delta_{ij},B_{ij}=b_i\delta_{ij}$,
we get the
 NC $(2+1)$-dimensional $N$-wave equation which is new:
\begin{eqnarray}
 \frac{\del U_{ij}}{\del t}=
 \lambda_{ij}\frac{\del U_{ij}}{\del x}
 -\mu_{ij}\frac{\del U_{ij}}{\del y}
 +\sum_{k=1}^{N}(\lambda_{ik}-\lambda_{kj})U_{ik}U_{kj},
\end{eqnarray}
where $\dis \lambda_{ij}:=(b_i-b_j)/(a_i-a_j),~
\mu_{ij}:=a_i\lambda_{ij}-b_i,~(i \neq j)$.
In the commutative limit, this equation is reduced to
the $(2+1)$-dimensional $N$-wave equation.
It is interesting to study whether NC $N$-wave equation
still admits a configuration of $N$ interacting waves
as commutative case \cite{AbHa}.

\item NC Davey-Stewartson equation

{}From now on, we assume $N=2$ and that 
all gauge fields depend on $t=z$ and $x=w+\tilde{w}$ only, 
which is another dimensional reduction.
The linear system (\ref{lin_asdym}) is reduced as follows:
\begin{eqnarray}
 \del_x \Psi=(A_w+\zeta A_{\tilde{z}})\star \Psi,~~~
 \del_t \Psi=(A_z+\zeta (A_w+A_{\tilde{w}})+\zeta^2 A_{\tilde{z}})
\star \Psi,
\end{eqnarray}
where the coefficient matrices depend on $x,y$ and $t$.
Now let us take $\Psi = \Phi e^{-\zeta y}$ and
further reduction condition:
$A_w=U+A\del_y,~A_z=V+\tilde{U}\del_y+A\del^2_y,
~A_{\tilde{z}}=A,~A_{\tilde{w}}=\tilde{U}-U+A\del_y$, 
we get a reduced linear system:
\begin{eqnarray}
 \del_x \Phi=(U +  A\del_y)\star \Phi,~~~
 \del_t \Phi=(V+\tilde{U}\del_y+A \del^2_y)\star \Phi.
\label{linear_ds}
\end{eqnarray}
By taking further reduction condition:
\begin{eqnarray}
A=\kappa\left(\begin{array}{cc}1&0\\0&-1\end{array}\right),~
U=\tilde{U}=\left(\begin{array}{cc}0&q\\r&0\end{array}\right),~
V=\frac{1}{2\kappa}\left(\begin{array}{cc}R_1&(\del_x+\kappa\del_y)q\\
-(\del_x-\kappa\del_y)r&R_2\end{array}\right),\nonumber
\end{eqnarray}
the compatibility condition 
yields a reduced ASDYM equation 
which is the NC Davey-Stewartson system:
\begin{eqnarray}
 2\kappa\dot{q}&=&(\del_x^2+\kappa^2\del_y^2)q
+R_1\star q-q\star R_2,\nn
 2\kappa\dot{r}&=&-(\del_x^2+\kappa^2\del_y^2)r
+R_2\star r-r\star R_1,
\end{eqnarray}
and 
\begin{eqnarray}
 (\del_x-\kappa\del_y)R_1&=&-(\del_x+\kappa\del_y)(q\star r),\nn
 (\del_x+\kappa\del_y)R_2&=&(\del_x-\kappa\del_y)(r\star q).
\end{eqnarray}
These are also new equations.
By taking a dimensional reduction $\del_y=0$,
this discussion is reduced to that of NC NLS equation
with $R_{1}=-q\star r,~R_{2}=r\star q,~q=\psi,~r=\tilde{\psi}$.

\item NC KP equation

NC KP equation can be derived from NC ASDYM equation
in similar discussion to NC DS equation.
Let us start from the reduced system (\ref{linear_ds}),
and take another further reduction
on the gauge fields as follows:
\begin{eqnarray}
A=\left(\begin{array}{cc}0&0\\1&0\end{array}\right),
U=\left(\begin{array}{cc}0&1\\u&0\end{array}\right),
\tilde{U}=\left(\begin{array}{cc}0&1\\u/2&0\end{array}\right),
V=\frac{1}{4}
\left(\begin{array}{cc} \alpha+u_x& -2u \\
 u_{xx}-2u\star u +u_y&\alpha-u_x
\end{array}\right).\nonumber
\end{eqnarray}
Then the compatibility condition gives rise 
to a NC reduced ASDYM equation:
\begin{eqnarray}
 \left(\begin{array}{cc} 
-\alpha_x-3u_y&~0\\
 4u_t-u_{xxx}+3u_x\star u
+3u\star u_x +\alpha_y +[u,\alpha]_\star
&~-\alpha_x-3u_y
\end{array}\right)=0.
\end{eqnarray}
Hence we get $\alpha=-3\del_{x}^{-1}u_y:=
-3\int^{x}u_y(x^\prime) dx^\prime$,
and the NC KP equation \cite{Kupershmidt,Paniak}:
\begin{eqnarray}
  u_t = \frac{1}{4} \left(u_{xxx}
  -3\left(u_x \star u + u \star u_x \right)
  +3 \del_{x}^{-1}u_{yy} +3[u,\del_{x}^{-1}u_y]_\star \right).
\end{eqnarray}
If we neglect $y$-dependence, the present discussion
gives rise to another reduction to NC KdV equation
from ASDYM equation with $G=SL(2,\R)$, 
which is reduced to discussion by Bakas and 
Depireux \cite{BaDe} in the commutative limit.

\end{itemize}

Reduction of NC ASDYM hierarchies 
to NC integrable hierarchies would be possible
in similar ways. The detailed discussion 
would be reported somewhere.

\section{Conclusion and Discussion}

In the present letter, we proved that
various NC integrable equations
can be derived from NC ASDYM equations in the split signature,
which include NC versions of 
KdV, NLS, $N$-wave, DS and KP equations.
$U(1)$ part of gauge groups
for the original Yang-Mills equations play crucial roles 
in NC extension of the Mason-Sparling's beautiful result.
Existence of these reductions guarantees
the lower-dimensional integrable equations actually have the
corresponding physical situations, such as, reduced D0-D4
D-brane systems. Then, Analysis of exact NC soliton solutions 
could be applied to that of  D-brane dynamics 
in the special reduced situations.

In reductions of NC ASDYM equations, we mainly should 
take the split signature.
NC Yang-Mills theories with the split signature
can be embedded \cite{LPS2} in
N=2 string theories \cite{OoVa}.  
Simple reductions of them and relation to
the string theories have been studied 
intensively \cite{LePo,Bieling,Wolf,IhUh,ChLe}.
This guarantees that NC integrable equations would 
have physical meanings and might lead to various successful 
applications to the corresponding D-brane
dynamics and so on. 

One of the next steps 
is to study supersymmetric extension (e.g \cite{Legare,NiRa})
of these reductions of NC ASDYM equations
in order to clarify what reductions 
are realized in the framework of D-brane systems
in N=2 string theories.
We can judge what reductions are better or worse
from the viewpoint of the number of the preserved
supersymmetry or stability of the corresponding
D-brane configurations.
This might exclude the special reductions 
to any conservation laws as we commented in the end of 
section 3.

The BPS equations in some D-brane configurations
would just correspond to NC integrable equations
and the soliton solutions correspond to 
lower-dimensional D-branes. 
In these situations, we can expect that 
similar applications to D-brane dynamics
would be possible, as in tachyon condensations. 
NC solitons are sometimes very easy to treat
in infinite noncommutativity limit.
The typical example is known as
the Gopakumar-Minwalla-Strominger (GMS) soliton \cite{GMS}.
GMS-like solitons would exist in our situations
in some limit and we could construct exact soliton solutions
and analyze energy densities of them and fluctuation spectrum
around them and so on. This would be a hint to reveal
the corresponding D-brane configurations which might 
be new BPS states.

\subsection*{Acknowledgments}

The author would like to thank 
H.~Kanno, O.~Lechtenfeld, 
L.~Mason, S.~Moriyama, F.~M\"uller-Hoissen,
A.~Popov and K.~Takasaki 
for fruitful discussions and useful comments.
He is also grateful to O.~Lechtenfeld,
F.~M\"uller-Hoissen and K.~Lee for financial support and
hospitality during stays  
at the Institute for Theoretical Physics, 
University of Hannover, 
and the Max Planck Institute for dynamics 
and self-organization, G\"ottingen,
and the Korea Institute for Advanced Study,
on March 2005. 
This work was partially supported 
by the Daiko Foundation (\#9095).



\end{document}